\documentclass[3p,times,procedia]{elsarticle}
\usepackage{nupha_ecrc}

\volume{00}

\firstpage{1}

\journalname{Nuclear Physics A}

\runauth{U. G\"ursoy et al.}

\jid{nupha}

\jnltitlelogo{Nuclear Physics A}

\usepackage{amssymb}
\usepackage{xcolor}

\usepackage[figuresright]{rotating}

\begin{document}

\begin{frontmatter}



\dochead{XXVIIIth International Conference on Ultrarelativistic Nucleus-Nucleus Collisions\\ (Quark Matter 2019)}

\title{Charge-dependent flow induced by electromagnetic fields\\ in heavy ion collisions}
\footnote{Presented by D. Kharzeev}


\author{U. G\"ursoy$^a$, D.E. Kharzeev$^{b,c,d}$, E. Marcus$^a$, K. Rajagopal$^e$ and C. Shen$^{f,d}$}
\address{$^a$ Institute for Theoretical Physics, Utrecht University, Leuvenlaan 4, 3584 CE Utrecht, The Netherlands}

\address{$^b$ Department of Physics and Astronomy, Stony Brook University, New York 11794, USA}
\address{$^c$ Physics Department, Brookhaven National Laboratory, Upton, NY 11973, USA}
\address{$^d$ RIKEN-BNL Research Center, Brookhaven National Laboratory, Upton, NY 11973, USA}


\address{$^e$ Center for Theoretical Physics, Massachusetts Institute of Technology, Cambridge, MA 02139}

\address{$^f$ Department of Physics and Astronomy, Wayne State University, Detroit, MI 48201, USA}

\begin{abstract}
The colliding heavy ions create extremely strong magnetic and electric fields that significantly affect the evolution of the produced quark-gluon plasma (QGP). The knowledge of these fields is essential for establishing the role of topological fluctuations in the QGP through the chiral magnetic effect and related anomaly-induced phenomena. In this talk, we describe our work on the evolution of the QGP in electric and magnetic fields in the framework of hydrodynamics supplemented, in a perturbative fashion, by the dynamical electromagnetism.  The evolution of the QGP fluid is described within the {\tt iEBE-VISHNU} framework. We find that the electromagnetically induced  currents result in a charge-odd directed flow
$\Delta v_1$ and a charge-odd $\Delta v_3$ flow both of which are odd in rapidity. While the predicted magnitude of these charge-odd flows agrees with the data from RHIC and LHC, the sign of the predicted asymmetry between the flows of positive and negative hadrons is opposite to the data. 
\end{abstract}


\end{frontmatter}


\section{Introduction}
\label{}
Non-central heavy ion collisions produce extremely strong magnetic fields~\cite{Kharzeev:2007jp,Skokov:2009qp,Tuchin:2010vs}. For such strong fields, electromagnetic interactions significantly affect the evolution of the produced quark-gluon plasma (QGP). In particular, the interplay between magnetic fields and chiral anomaly has been predicted to lead to interesting phenomena including the chiral magnetic effect~\cite{kharzeev2006parity,Kharzeev:2007jp,Fukushima:2008xe}
and the chiral magnetic wave~\cite{Kharzeev:2010gd,Burnier:2011bf}. These quantum chiral effects, if observed, would reveal the existence of topological charge fluctuations in the QGP that are similar to the topological fluctuations in the primordial electroweak plasma responsible for the baryon asymmetry of the Universe. 

This motivation drives the need for establishing the presence and the magnitude of electromagnetic fields in the QGP. Here we report on the results of our studies \cite{Gursoy:2014aka, gursoy2018charge}  of the effects of electromagnetic fields on the collective flow of charged hadrons. 
Our treatment of electromagnetic effects is perturbative: we solve the Maxwell equations to evaluate the currents induced by electromagnetism on top of the conventional hydrodynamical evolution (no back-reaction is included). 

\section{Electromagnetic fields in heavy ion collisions: a qualitative discussion}
\begin{figure}[t]
 \begin{center}
\includegraphics[scale=0.35]{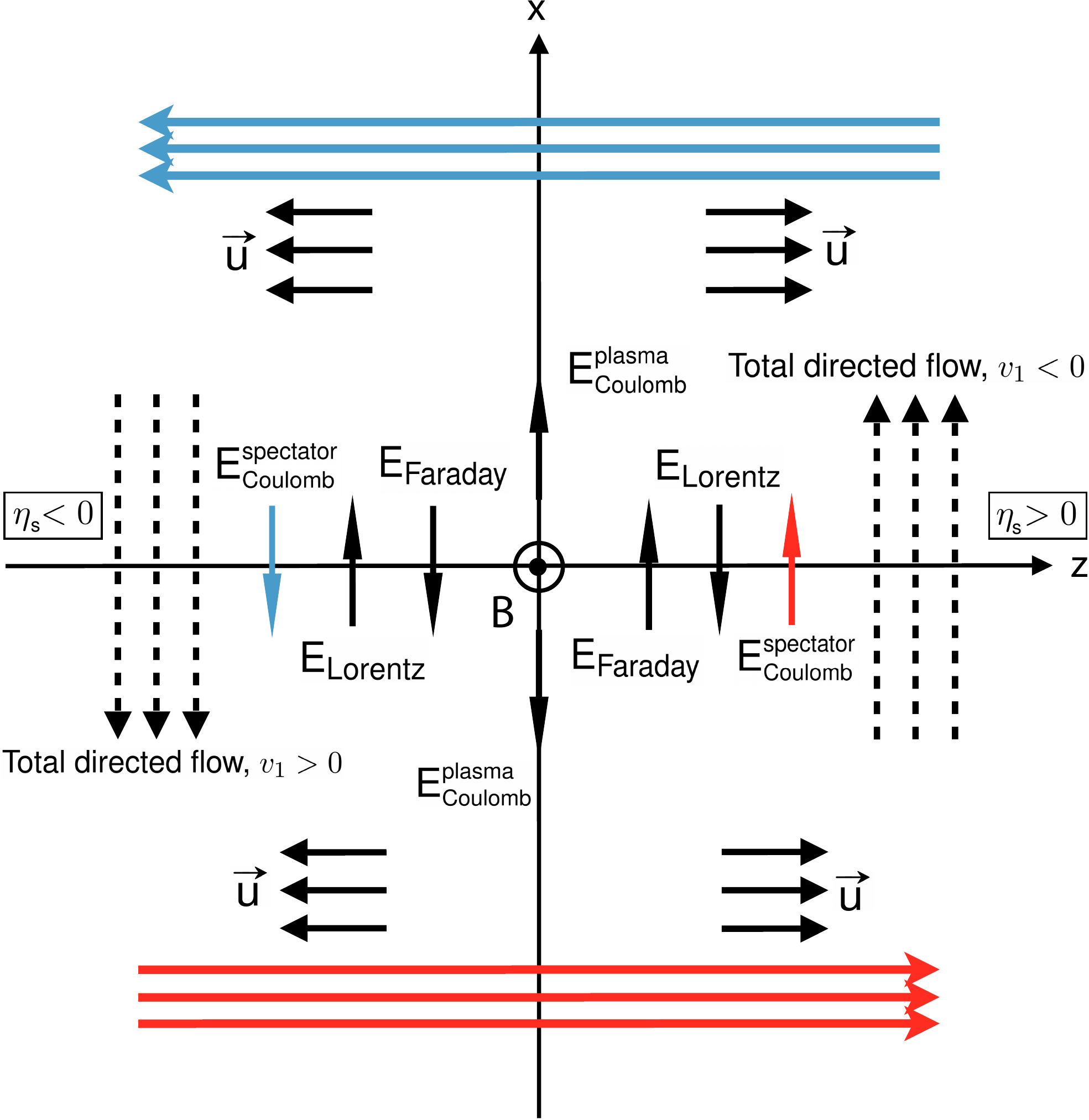}
 \end{center}
 \vspace{-0.3in}
 \caption{Schematic illustration of how the electromagnetic fields in a heavy ion collision result
in a directed flow of electric charge, $\Delta v_1$, see text for explanation. From \cite{gursoy2018charge}.}
\label{figschema}
\end{figure}

\begin{figure}[t!]
 \begin{center}
\includegraphics[scale=0.28]{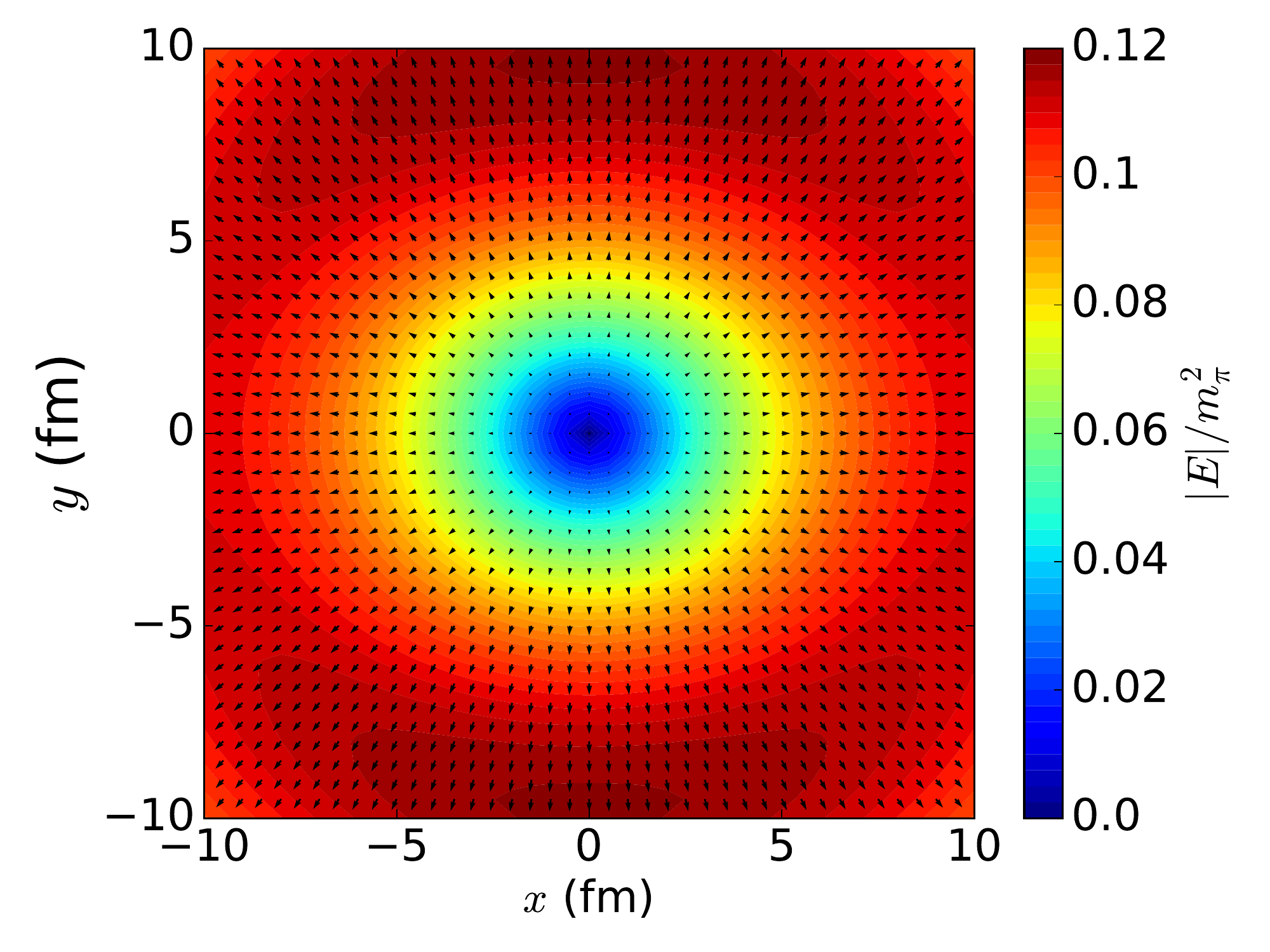}
\includegraphics[scale=0.29]{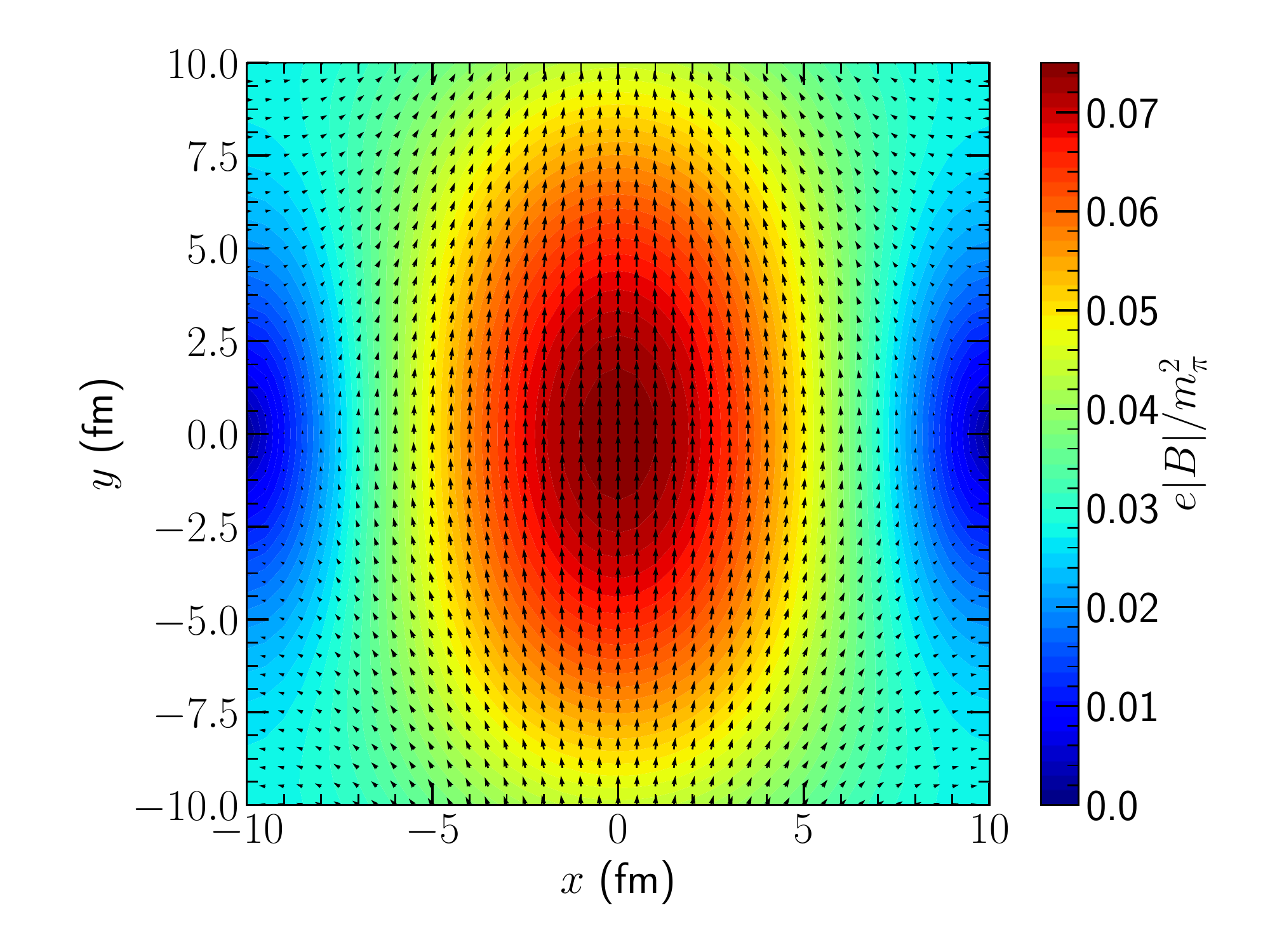}
 \end{center}
 \caption[]{The electric (left) and magnetic (right) fields in the transverse plane at $z=0$  in the lab frame 
 at a proper time $\tau=1$~fm/c after a Pb+Pb collision
 with 20-30\% centrality (impact parameters in the range 6.24~fm~$<b<$~9.05~fm) and with a collision energy $\sqrt{s} =2.76$~ATeV.   From \cite{gursoy2018charge}. }
\label{figEB}
\end{figure}

As illustrated in Fig.~\ref{figschema}, there are three distinct electromagnetic effects 
on charged components of the fluid, resulting in a sideways current:
\begin{enumerate}
\item {\it Faraday:} as the magnetic field decreases in time, Faraday's law implies the induction of an electric field and, since the QGP possesses mobile charges, an electric current. We denote this electric field by $\vec E_F$.  Since $\vec E_F$ curls around the (decreasing in time) $\vec B$ that points in the $y$-direction, the sideways component of $E_F$ points in opposite directions at opposite rapidity, see Fig.~\ref{figschema}.
\item {\it Lorentz:} since the QGP fluid exhibits a strong longitudinal flow velocity $\vec{v}_\mathrm{flow}$ denoted by $\vec u$ in Fig.~\ref{figschema} pointing along the beam direction (and hence perpendicular to $\vec B$), the Lorentz force exerts a sideways push on charged particles in opposite directions at opposite rapidity. Equivalently, upon boosting to the local fluid rest frame, the lab frame $\vec B$ yields a fluid frame $\vec E$ in which the effects on the charged components of the fluid are equivalent to the effects of the Lorentz force in the lab frame. We denote this electric field by $\vec E_L$.  Both $\vec E_F$ and $\vec E_L$ originate from the magnetic field.
\item {\it Coulomb:} The positive spectators exert an electric force on the charged plasma produced in the collision, which again points in opposite directions at opposite rapidity; the corresponding 
electric field is denoted by $\vec E_C$. 
\end{enumerate}
All three of these electric fields --- and the electric currents induced by them --- have opposite directions at positive and negative rapidity. As shown in Fig.~\ref{figschema}, 
$\vec E_F$ and  $\vec E_C$ have the same sign, whereas  $\vec E_L$ opposes them. Therefore, 
 the sign of the total rapidity-odd, charge-odd, directed flow charge dependence $\Delta v_1$ that results 
 from these electric fields
 depends on whether  $\vec E_F + \vec E_C$ or $\vec E_L$ is dominant. The relative magnitude of these fields is quite sensitive to the space-time evolution of the fluid.

\section{Electromagnetic fields in heavy ion collisions: numerical results}

Our numerical results were obtained, as explained in \cite{gursoy2018charge}, by describing the evolution of the QGP fluid within the {\tt iEBE-VISHNU} framework \cite{Shen:2014vra}, and by solving Maxwell equations on top of this hydrodynamical background.  The 
hydrodynamic calculation 
assumes longitudinal boost-invariance and begins at $\tau_0 = 0.4$ fm/$c$. We evolve the relativistic viscous
hydrodynamic equations for a fluid with an equation of state based upon lattice QCD calculations \cite{Huovinen:2009yb}. The electromagnetic fields generated by the charges and currents evolve according to the Maxwell equations, in which we assume a constant conductivity. 

\begin{figure*}[t!]
  \includegraphics[width=0.6\linewidth]{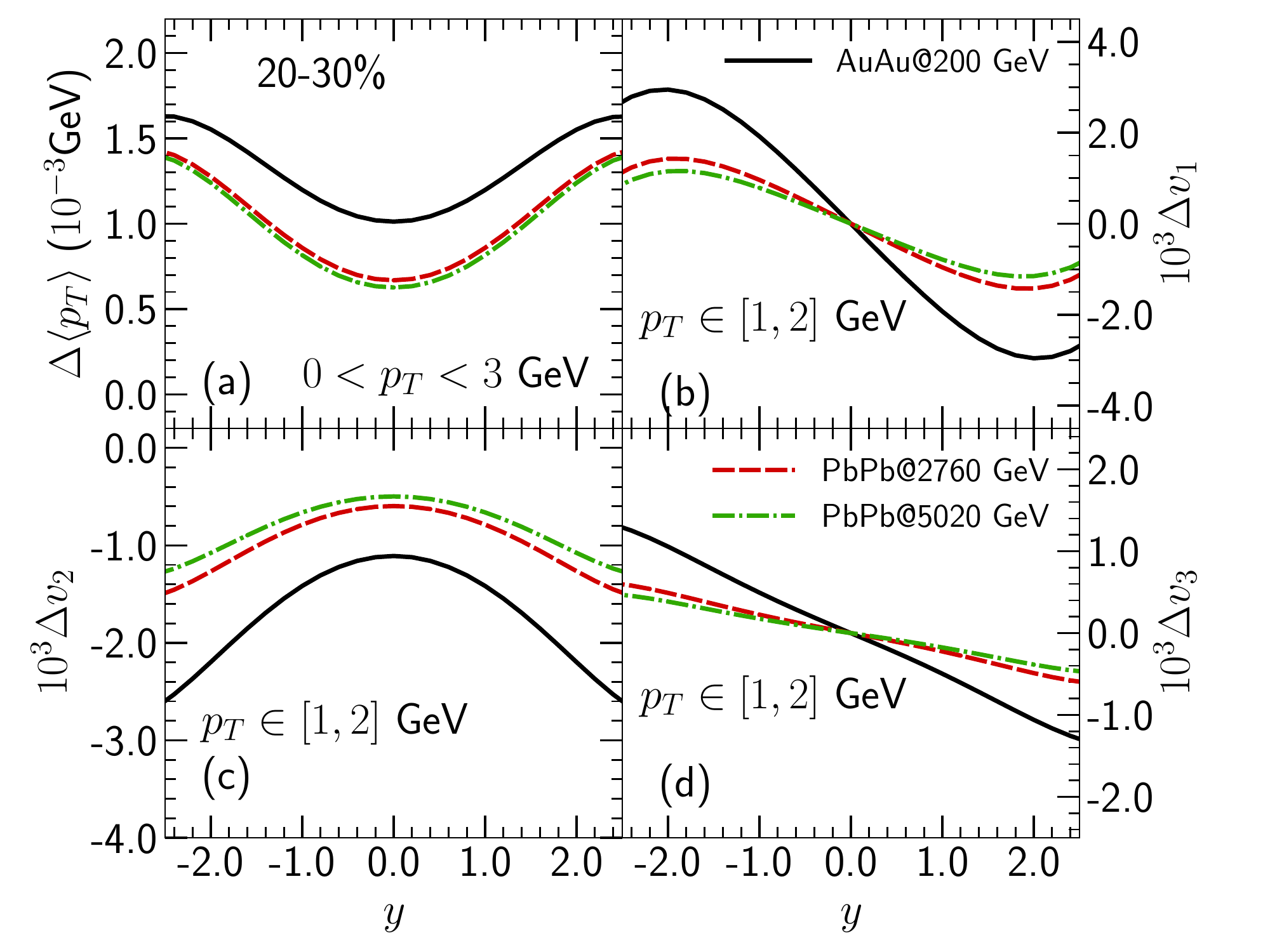}   
  \caption{The collision energy dependence of the electromagnetically induced charge-odd contributions to flow observables. The difference of particle mean $p_T$ and $v_n$ between $\pi^+$ and $\pi^-$ are plotted as a function of particle rapidity for collisions at the top RHIC energy of 200 GeV and at two LHC collision energies; from \cite{gursoy2018charge}..}
  \label{fig5}
\end{figure*}

Our calculation yields the momentum distribution for hadrons with different charge, from which we  evaluate  the difference between the flow for positively and negatively charged hadrons 
$\Delta v_n \equiv v_n(h^+) - v_n(h^-)$ 
shown in Fig.~\ref{fig5} for the top RHIC and LHC energies.

\section{Conclusions}

The data from RHIC \cite{Adam:2019wnk} and LHC \cite{Acharya:2019ijj} show that the observed $\Delta v_1$ is similar in magnitude to the one that we have predicted, but has an {\it opposite sign}. This 
indicates that the electromagnetic fields of expected magnitude are indeed present in the QGP, but our treatment of magnetohydrodynamics of the plasma has to be improved. This can be done along two directions: i) improve the computation within the perturbative framework, e.g. by including the time-dependent conductivity; and ii) perform a full magnetohydrodynamical (MHD) simulation. The recent pilot MHD simulation \cite{Inghirami:2019mkc} however indicates the same sign of $\Delta v_1$ as our perturbative result. Clearly, more work is needed to improve our understanding of this important issue.

\section{Acknowledgements}
This work was supported in part by the Netherlands Organisation for Scientific Research (NWO) under VIDI grant 680-47-518, the Delta Institute for Theoretical Physics (D-ITP) funded by the Dutch Ministry of Education, Culture and Science (OCW), the Scientific and Technological Research Council of Turkey (TUBITAK), the Office of Nuclear Physics of the U.S.~Department of Energy under Contract Numbers  DE-SC0011090, DE-FG-88ER40388 and DE-AC02-98CH10886, and the Natural Sciences and Engineering Research Council of Canada. KR gratefully acknowledges the hospitality of the CERN Theory Group. CS gratefully acknowledges a Goldhaber Distinguished Fellowship from Brookhaven Science Associates. Computations were made in part on the supercomputer Guillimin from McGill University, managed by Calcul Qu\'ebec and Compute Canada. The operation of this supercomputer is funded by the Canada Foundation for Innovation (CFI), NanoQu\'ebec, RMGA and the Fonds de recherche du Qu\'ebec -- Nature et technologies (FRQ-NT). UG is grateful for the hospitality of the Bo\~gazi\c ci University and the Mimar Sinan University in Istanbul. We gratefully acknowledge helpful discussions with Gang Chen, Ulrich Heinz, Jacopo Margutti, Raimond Snellings, Sergey Voloshin and Fuqiang Wang. 





\bibliographystyle{elsarticle-num}
\bibliography{references}

\begin{thebibliography}{10}
\expandafter\ifx\csname url\endcsname\relax
  \def\url#1{\texttt{#1}}\fi
\expandafter\ifx\csname urlprefix\endcsname\relax\def\urlprefix{URL }\fi
\expandafter\ifx\csname href\endcsname\relax
  \def\href#1#2{#2} \def\path#1{#1}\fi

\bibitem{Kharzeev:2007jp}
D.~E. Kharzeev, L.~D. McLerran, H.~J. Warringa, {The Effects of topological
  charge change in heavy ion collisions: 'Event by event P and CP violation'},
  Nucl. Phys. A803 (2008) 227--253.
\newblock \href {http://arxiv.org/abs/0711.0950} {\path{arXiv:0711.0950}},
  \href {http://dx.doi.org/10.1016/j.nuclphysa.2008.02.298}
  {\path{doi:10.1016/j.nuclphysa.2008.02.298}}.

\bibitem{Skokov:2009qp}
V.~Skokov, A.~{\relax Yu}. Illarionov, V.~Toneev, {Estimate of the magnetic
  field strength in heavy-ion collisions}, Int. J. Mod. Phys. A24 (2009)
  5925--5932.
\newblock \href {http://arxiv.org/abs/0907.1396} {\path{arXiv:0907.1396}},
  \href {http://dx.doi.org/10.1142/S0217751X09047570}
  {\path{doi:10.1142/S0217751X09047570}}.

\bibitem{Tuchin:2010vs}
K.~Tuchin, {Synchrotron radiation by fast fermions in heavy-ion collisions},
  Phys. Rev. C82 (2010) 034904, [Erratum: Phys. Rev.C83,039903(2011)].
\newblock \href {http://arxiv.org/abs/1006.3051} {\path{arXiv:1006.3051}},
  \href {http://dx.doi.org/10.1103/PhysRevC.83.039903,
  10.1103/PhysRevC.82.034904} {\path{doi:10.1103/PhysRevC.83.039903,
  10.1103/PhysRevC.82.034904}}.

\bibitem{kharzeev2006parity}
D.~Kharzeev, Parity violation in hot qcd: Why it can happen, and how to look
  for it, Physics Letters B 633~(2-3) (2006) 260--264.

\bibitem{Fukushima:2008xe}
K.~Fukushima, D.~E. Kharzeev, H.~J. Warringa, {The Chiral Magnetic Effect},
  Phys. Rev. D78 (2008) 074033.
\newblock \href {http://arxiv.org/abs/0808.3382} {\path{arXiv:0808.3382}},
  \href {http://dx.doi.org/10.1103/PhysRevD.78.074033}
  {\path{doi:10.1103/PhysRevD.78.074033}}.

\bibitem{Kharzeev:2010gd}
D.~E. Kharzeev, H.-U. Yee, {Chiral Magnetic Wave}, Phys. Rev. D83 (2011)
  085007.
\newblock \href {http://arxiv.org/abs/1012.6026} {\path{arXiv:1012.6026}},
  \href {http://dx.doi.org/10.1103/PhysRevD.83.085007}
  {\path{doi:10.1103/PhysRevD.83.085007}}.

\bibitem{Burnier:2011bf}
Y.~Burnier, D.~E. Kharzeev, J.~Liao, H.-U. Yee, {Chiral magnetic wave at finite
  baryon density and the electric quadrupole moment of quark-gluon plasma in
  heavy ion collisions}, Phys. Rev. Lett. 107 (2011) 052303.
\newblock \href {http://arxiv.org/abs/1103.1307} {\path{arXiv:1103.1307}},
  \href {http://dx.doi.org/10.1103/PhysRevLett.107.052303}
  {\path{doi:10.1103/PhysRevLett.107.052303}}.

\bibitem{Gursoy:2014aka}
U.~Gursoy, D.~Kharzeev, K.~Rajagopal, {Magnetohydrodynamics, charged currents
  and directed flow in heavy ion collisions}, Phys. Rev. C89~(5) (2014) 054905.
\newblock \href {http://arxiv.org/abs/1401.3805} {\path{arXiv:1401.3805}},
  \href {http://dx.doi.org/10.1103/PhysRevC.89.054905}
  {\path{doi:10.1103/PhysRevC.89.054905}}.

\bibitem{gursoy2018charge}
U.~G{\"u}rsoy, D.~Kharzeev, E.~Marcus, K.~Rajagopal, C.~Shen, Charge-dependent
  flow induced by magnetic and electric fields in heavy ion collisions,
  Physical Review C 98~(5) (2018) 055201.

\bibitem{Shen:2014vra}
C.~Shen, Z.~Qiu, H.~Song, J.~Bernhard, S.~Bass, U.~Heinz, {The iEBE-VISHNU code
  package for relativistic heavy-ion collisions}, Comput. Phys. Commun. 199
  (2016) 61--85.
\newblock \href {http://arxiv.org/abs/1409.8164} {\path{arXiv:1409.8164}},
  \href {http://dx.doi.org/10.1016/j.cpc.2015.08.039}
  {\path{doi:10.1016/j.cpc.2015.08.039}}.

\bibitem{Huovinen:2009yb}
P.~Huovinen, P.~Petreczky, {QCD Equation of State and Hadron Resonance Gas},
  Nucl. Phys. A837 (2010) 26--53.
\newblock \href {http://arxiv.org/abs/0912.2541} {\path{arXiv:0912.2541}},
  \href {http://dx.doi.org/10.1016/j.nuclphysa.2010.02.015}
  {\path{doi:10.1016/j.nuclphysa.2010.02.015}}.

\bibitem{Adam:2019wnk}
J.~Adam, et~al., {First Observation of the Directed Flow of $D^{0}$ and
  $\overline{D^0}$ in Au+Au Collisions at $\sqrt{s_{\rm NN}}$ = 200 GeV}, Phys.
  Rev. Lett. 123~(16) (2019) 162301.
\newblock \href {http://arxiv.org/abs/1905.02052} {\path{arXiv:1905.02052}},
  \href {http://dx.doi.org/10.1103/PhysRevLett.123.162301}
  {\path{doi:10.1103/PhysRevLett.123.162301}}.

\bibitem{Acharya:2019ijj}
S.~Acharya, et~al., {Probing the effects of strong electromagnetic fields with
  charge-dependent directed flow in Pb-Pb collisions at the LHC}\href
  {http://arxiv.org/abs/1910.14406} {\path{arXiv:1910.14406}}.

\bibitem{Inghirami:2019mkc}
G.~Inghirami, M.~Mace, Y.~Hirono, L.~Del~Zanna, D.~E. Kharzeev, M.~Bleicher,
  {Magnetic fields in heavy ion collisions: flow and charge transport}\href
  {http://arxiv.org/abs/1908.07605} {\path{arXiv:1908.07605}}.

\end{thebibliography}







\end{document}